\pdfoutput=1
\documentclass[aps,prd,amsmath,floats,floatfix, twocolumn,
superscriptaddress,nofootinbib,showpacs,longbibliography]{revtex4-1}

\usepackage[T1]{fontenc}
\usepackage[utf8]{inputenc}
\usepackage{lmodern}
\usepackage{color,soul}
\usepackage{verbatim}

\usepackage[dvipsnames, usenames]{xcolor}
\definecolor{linkcolor}{rgb}{0.0,0.3,0.5}
\usepackage[hypertexnames=false, unicode, colorlinks=true, linkcolor=linkcolor,
citecolor=linkcolor, filecolor=linkcolor,urlcolor=linkcolor,
pdfusetitle]{hyperref}

\usepackage[all]{hypcap}
\usepackage{graphicx}
\usepackage{xspace}
\usepackage{amssymb}
\usepackage[normalem]{ulem} %for \sout
\usepackage{bm} % boldmath

\usepackage{microtype}

\usepackage[english]{babel}
\usepackage{blindtext}

\usepackage{xcolor}   
\newcommand{\rev}{}

\graphicspath{%
  {figs/}%
}

\DeclareMathAlphabet{\mathpzc}{OT1}{pzc}{m}{it}

\DeclareMathOperator\supp{supp}

\begin{document}

\title{\rev{Looping back to the past through free fall in a controlled warp drive spacetime}}

\newcommand{\UQ}{\affiliation{Centre for Quantum Computation and Communication Technology, School of Mathematics and Physics, The University of Queensland, St. Lucia, Queensland, 4072, Australia}}

\author{Achintya Sajeendran}
\email{a.sajeendran@uq.edu.au}

\author{Timothy C. Ralph}
\email{ralph@physics.uq.edu.au}

\UQ 
% Because hyperref only gets the *last* author, we need to be explicit.
\hypersetup{pdfauthor={Sajeendran et al.}}

\date{\today}

%==========================================================================
\begin{abstract}
   \rev{We present a modification to a 'rotating' version of the dynamical Alcubierre spacetime, which was previously shown to permit closed timelike curves. We find that if the effective rotation rate is made dependent on the spacetime coordinates within the bubble, a class of closed timelike curves are promoted to spatially circular geodesics. These paths provide a simple model of a free particle interacting with a CTC for a finite proper time interval, entering and exiting in flat spacetime. Despite the questionable physical realisability of our metric, we suggest that it may provide a useful background for future theoretical studies of classical and quantum models of time travel in a general relativistic setting.
}
\end{abstract}

\maketitle

%==========================================================================
\section{Introduction}
\label{sec:introduction}
Since G\"{o}del's seminal paper \cite{godel1949example}, it has been known that certain solutions of the Einstein field equations in general relativity (GR) permit closed timelike curves (CTCs), paths that allow backwards time travel. Strictly speaking, CTCs are worldlines that are closed in both space and time, meaning that a particle that enters one when $t=0$ also exits it when $t=0$ at the same spatial point. 

\rev{Broadly speaking, there are two different classes of spacetimes permitting time travel that have been widely explored in the literature: exact solutions of the Einstein field equations \cite{godel1949example,van1938ix,tipler1974rotating,misner1963flatter,misner1967taub,gott1991closed,mallary2018closed}, and ad hoc designed metrics specifically constructed to permit time travel or superluminal travel \cite{ori1993must,ori1994causality,ori2005class,soen1996improved,tippett2017traversable,morris1988wormholes,alcubierre1994warp,everett1996warp,fermi2018time,ralph2020spinning,shoshany2023warp,maclaurin2024falling}.\footnote{Metrics constructed to allow (effective) superluminal travel can typically be modified to allow for time travel. The spacetime considered in this paper is one such metric, derived as a modification of the Alcubierre `warp drive' metric \cite{alcubierre1994warp}. The traversable wormhole metric \cite{morris1988wormholes} is another well-known example.} For spacetimes of the second class, the metric is first constructed without consideration of the physical matter distribution required to produce it, and the stress-energy tensor is subsequently `derived' from the right-hand side of the Einstein equations. Such constructions typically exhibit exotic, seemingly unphysical features, such as violations of the energy conditions. It is generally the case that these issues are either ignored altogether or treated separately. While such ad hoc spacetimes cannot generally be considered as solutions of the classical Einstein equations, they cannot be completely ruled out as unphysical, as the required negative energy densities may be achievable through quantum effects \cite{casimir1948influence,ford1978quantum,pfenning1998quantum,funai2017engineering}.}

\rev{Whilst true that CTCs can be found in a number of spacetimes, there is no compelling reason to believe that these spacetimes model any patch of our Universe or could somehow be realised. Indeed, Hawking's chronology protection conjecture \cite{hawking1992chronology} seems to suggest that quantum effects may actually prevent the formation of CTCs entirely. However, this conjecture is based on a semiclassical treatment, in which matter fields are treated as quantum, while gravity is treated classically. This approximation is known as \emph{semiclassical gravity}, as it neglects backreaction effects due to quantum fluctuations of the field. It has been suggested that this limitation of semiclassical gravity can lead to anomalous predictions, such as the cosmological constant problem \cite{cree2018can,wang2020vacuum,wang2020reformulation}. Therefore, it may not actually be true that quantum effects prevent the formation of CTCs, and a complete theory of quantum gravity may be required to seriously address this question.}

On the other hand, further research into quantum mechanics near CTCs may lead to useful insight into quantum gravity. To this end, several (non-linear and non-unitary) extensions of quantum mechanics that remain self-consistent near CTCs have been proposed and thoroughly investigated in the literature \cite{deutsch1991quantum,politzer1994path,lloyd2011quantum,lloyd2011closed,ralph2009quantum}. \rev{However, each of these models feature different departures from standard quantum mechanics \cite{ahn2013quantum,bub2014quantum,brun2009localized,brun2012perfect,ralph2011problems,bishop2022billiard}, and it seems that assessing these models purely at the level of mathematical self-consistency is insufficient for determining which of them are the most physically viable.}

In quantum models of CTCs, the CTC itself is modelled as \rev{being due to a} traversable wormhole \cite{morris1988wormholes} embedded in flat spacetime. These approaches are idealised in the sense that they neglect the spacetime geometry separating the Minkowski spacetime region and the two wormhole mouths, assuming the transition between the two geometries to be instantaneous. \rev{This is because the wormhole mouths are treated as points in Minkowski space. More problematically, the complicated geometry of the wormhole throat is completely ignored.} Thus, the actual gravitational field (or curvature) required for the formation of such a CTC is abstracted away. Given the complicated structure of most of the spacetimes that permit CTCs, there is strong motivation for making such a simplification. Furthermore, in a sense, it is more general to consider a CTC without explicit reference to an actual spacetime background, as one can argue that backwards time travel is the only feature that needs to be captured in the model. On the other hand, introducing gravity into the picture may result in additional physical restrictions, which might serve as a useful benchmark for the existing quantum models of time travel. The present work aims to make progress within this largely unexplored area, using a relatively simple CTC metric proposed by Ralph and Chang \cite{ralph2020spinning}. In this paper, we modify the Ralph and Chang metric to allow for simple timelike geodesics that are also CTCs, with the property that particles enter and exit the CTC in chronology-respecting (Minkowski) space. Spacetimes with this property would be ideal for studying quantum effects near CTCs, \rev{or even testing existing classical models} \cite{friedman1990cauchy,echeverria1991billiard,tobar2020reversible}.\footnote{Other examples of such spacetimes have been studied by Fermi and Pizzocchero \cite{fermi2018time}, and recently by Shoshany and Snodgrass \cite{shoshany2023warp} in a more precise formulation of Everett's spacetime \cite{everett1996warp} (see Section \ref{sec:rotatingalc}).} 

Throughout this work, we employ natural units in which $c=1$. For heavy computations involving tensors, we have used the \emph{OGRe} package \cite{Shoshany2021_OGRe} in \emph{Mathematica}.

%==========================================================================
\section{The Alcubierre warp drive}
\label{sec:rotalc}

Inspired by the phenomenon of the expanding universe, Alcubierre \cite{alcubierre1994warp} proposed an \rev{ad hoc spacetime metric, constructed using the ADM formalism,} allowing for (apparent) superluminal travel. Specifically, the key feature of the spacetime is the Alcubierre \emph{bubble}; a localised pocket of space such that spacetime expands behind it, while contracting in front of it. \rev{Typically the bubble is taken to be compactly supported, so that the spacetime outside of it is flat.} As such, a particle inside the bubble could appear to travel at faster than the speed of light according to a Minkowski observer outside the bubble. Note that this does not violate causality, as the particle itself is not moving through space at a locally superluminal velocity. Rather, it is the stretching of spacetime itself near the particle that makes it appear to be moving superluminally according to the observer. This is similar to how during the inflationary phase of the Universe, two comoving observers would appear to be moving away from each other at superluminal speed, as Alcubierre argued in his original paper \cite{alcubierre1994warp}.

The original Alcubierre metric, colloquially known as the \emph{warp drive}, takes on the form \cite{alcubierre1994warp},
\begin{equation}\label{eq:alcubierre}
    ds^2=-(1-f^2 v_s^2)\:dt^2-2fv_s\: dx \: dt+dx^2+dy^2+dz^2
\end{equation}
Here, $f=f(t,x,y,z)$ is a function that determines the shape of the bubble; it is equal to $1$ near the centre of the bubble, and smoothly decreases to $0$ away from the bubble so that the spacetime is asymptotically flat. More precisely, $f$ is usually taken to be compactly supported so that the bubble is a localised region, outside of which the spacetime is completely flat. \rev{The warp velocity, $v_s=v_s(t)$, is effectively the velocity of the bubble. This is in the sense that a test particle inside the bubble travelling at velocity $v_s$ according to a Minkowksi observer suffers no time dilation relative to the observer; that is, the proper time of the particle is equal to the Minkowski time, $t$, in a certain inertial reference frame.} Furthermore, if we set $v_s>1$, such a particle would appear superluminal.

One significant drawback of the Alcubierre spacetime is that its stress-energy tensor violates all classical energy conditions \cite{alcubierre1994warp}, meaning that the energy density around the bubble is not positive definite for all observers. In particular, the formation of the contracting region in front of the bubble requires a large negative energy density. Thus, at the classical level, such a spacetime would need to be sourced by exotic matter, which has not been found in nature. On the other hand, quantum fields immediately violate the pointwise energy conditions due to quantum fluctuations \cite{carroll2019spacetime,kontou2020energy}, \rev{and negative energy densities can be formed for example through the Casimir effect \cite{casimir1948influence} or quantum energy teleportation \cite{hotta2009quantum,funai2017engineering}.}

\rev{The ability of quantum fields to violate the pointwise energy conditions motivated the formulation of \emph{quantum energy inequalities} (QEIs), which impose restrictions on the negative energy densities that can be generated by any configuration involving relativistic quantum fields \cite{ford1978quantum,pfenning1998quantum} (see Ref. \cite{kontou2020energy} for a review).

Pfenning \cite{pfenning1998quantum} originally showed that applying QEIs to the Alcubierre warp drive leads to constraints on the negative energy region of the bubble, implying that the total negative energy needed for the formation of a warp drive was several times greater than the total mass contained in the observable Universe. In subsequent years, it was argued that this huge negative energy requirement could be dramatically reduced to the order of a few milligrams by modifying the structure of the bubble \cite{van1999awarp,krasnikov2003quantum}. One could therefore argue that the Alcubierre metric could be sourced by quantum matter fields. However, this line of reasoning neglects the emission of Hawking radiation within the bubble due to horizons at the front and back walls. It was first shown by Hiscock \cite{hiscock1997quantum} that an eternal superluminal Alcubierre bubble (one of constant warp velocity) would be rendered unstable in the presence of quantum fields, as the Hawking particles would effectively build up at the front wall of the bubble, resulting in a semiclassical backreaction that would likely destroy it. This was later generalised to dynamical (accelerating) warp drives by Finazzi \emph{et al.} \cite{finazzi2009semiclassical}. Therefore, it would seem that the very quantum fields that could be used to create the warp drive would also destroy it. It should be noted that these treatments neglect fluctuations of the stress-energy operator of the quantum fields, and there is still a possibility that quantum gravity may allow for the formation of warp drives. However, at the time of writing of this paper, we believe this argument is mere speculation.} For our purposes, we shall not discuss this issue any further. Instead, we will proceed by assuming that an Alcubierre bubble could somehow be formed and derive the consequences that follow, in a similar fashion to Alcubierre's original paper \cite{alcubierre1994warp}.

%--------------------------------------------------------------------------

%==========================================================================
\section{A rotating Alcubierre metric}
\label{sec:rotatingalc}

While the Alcubierre solution itself is globally hyperbolic, Alcubierre \cite{alcubierre1994warp} suggested that it might be possible to modify it such that the resulting spacetime permits CTCs. This seems intuitive, as protocols that allow for superluminal travel can generally be used to create a time machine with a slight modification. Indeed, Everett \cite{everett1996warp} found that a spacetime consisting of two Alcubierre bubbles, as defined in inertial frames related by a Lorentz boost, can allow for time travel.\footnote{Recently, a more precise formulation of Everett's spacetime has been derived in Ref. \cite{shoshany2023warp}.} More recently, Ralph and Chang \cite{ralph2020spinning} have proposed a different construction---namely, an Alcubierre bubble in a `rotating spacetime'---that permits CTCs in certain regimes.

\rev{Expressed in cylindrical coordinates}\footnote{\rev{That is, coordinates in which the line element outside the bubble is $$ds^2=-dt^2+r^2 d\phi^2+dr^2+dz^2.$$ This is simply the line element of Minkowski spacetime expressed in cylindrical coordinates.}} of the \rev{\emph{lab frame}}, the line element for the rotating Alcubierre spacetime takes on the form,
\rev{\begin{equation}\label{eq:global metric}
\begin{aligned}
ds^2 = & -\frac{1 - (\omega r + f v_s)^2}{1 - \omega^2 r^2} dt^2 \\
& - \frac{2\left(f v_s + \omega r f^2 v_s^2 + \omega^2 r^2 f v_s\right)}{1 - \omega^2 r^2} r d\phi dt \\
& + \frac{\left(1 + f v_s \omega r\right)^2 - \omega^2 r^2}{1 - \omega^2 r^2} r^2 d\phi^2 + dr^2 + dz^2.
\end{aligned}
\end{equation}}
Here, $f$ and $v_s$ retain the same physical interpretations they had in the original metric \eqref{eq:alcubierre}. For $\omega=0$, this metric describes an Alcubierre bubble moving circularly, i.e. tangentially in the $\phi$ direction. \rev{The rotating Alcubierre metric can then be loosely visualised as this setup combined with a rotating platform (as explained in Refs. \cite{ralph2020spinning,maclaurin2024falling}), or alternatively, an Alcubierre bubble `orbiting' a Kerr black hole. However, upon closer inspection, the metric presented here does not correspond to a physically rotating object, as the effect of the `rotation' is confined to the inside of the bubble. To see this, note that $f$ vanishes outside the bubble, by definition. Substituting $f=0$ into the metric yields the Minkowski metric. Therefore, there is no global frame dragging effect, implying that there is no physical rotation. There is only a local frame dragging effect inside the bubble, which can intuitively be thought of as originating from a row of massive conveyor belt-like objects inside the bubble,\footnote{\rev{Think of a conveyor belt situated at each radial coordinate $r$ lying inside the bubble.}} or the flow of some fluid that is confined to the bubble, with velocity $\omega r$, rather than a rotating platform. For brevity, we will hereafter refer to this confined frame dragging effect as a rotation, recognising that this term is not entirely accurate in describing the global geometry of the spacetime.}

Consider now a pointlike particle in the centre of the bubble at a constant radius $r=r_s$, with the bubble being localised around this radius. Ralph and Chang \cite{ralph2020spinning} then transform to the \emph{local frame} $(t',\phi',r,z)$, by making a tangential Lorentz boost by velocity $\omega r_s$, and evaluating the transformed metric near $r=r_s$. Physically, these are the coordinates of an observer undergoing circular motion at angular velocity $\omega$ near the radius at which the bubble resides. The line element in the local frame reduces to
\begin{align}
    ds^2&=-(1-f^2 v_s^2)\:dt'^2-2fv_s r_s\: d\phi'\: dt' \nonumber\\ &\hspace{0.4cm}+r_s^2\: d\phi'^2+dr^2+dz^2,
\end{align}
which simply describes an Alcubierre bubble in circular motion. We stress that this form of the metric is only valid locally, at $r=r_s$. \rev{One can interpret this transformation as locally (only at $r=r_s$) undoing the `rotational' dragging effect inside the bubble.}

The four-velocity of a massive point particle that stays inside the bubble can be easily determined in the local frame as 
\begin{equation}\label{eq:four vel local}
    u^{\mu'}=\left( \frac{dt'}{d\tau},\frac{d\phi'}{d\tau},\frac{dr}{d\tau},\frac{dz}{d\tau} \right)=\left( 1,\frac{fv_s}{r_s},0,0 \right).
\end{equation}
\rev{This can then be transformed into the lab frame, giving the four-velocity \cite{ralph2020spinning}
\begin{equation}\label{eq:four vel global}
    u^\mu=\left( \frac{1+v_s\omega r_s}{\sqrt{1-\omega^2 r_s^2}}, \frac{v_s+\omega r_s}{r_s\sqrt{1-\omega^2 r_s^2}},0,0 \right).
\end{equation}}
Here, the assumption is made that the particle remains at the centre of the bubble, where $f=1$. Notice from Eq. \eqref{eq:four vel global} that if we take $\omega<0$ and $v_s>0$ such that $v_s\omega r<-1$, then $u^t=\frac{dt}{d\tau}<0$. Physically, this means that the worldline of the particle is past-pointing in this regime, thus allowing time travel. In particular, consider a dynamical warp drive that accelerates from $v_s=0$ up to some $v_s=v_c>1$ such that $v_s\omega r<-1$, and eventually decelerates back to $v_s=0$. Ralph and Chang \cite{ralph2020spinning} showed that this simple circular trajectory becomes a CTC if $v_s=v_c$ for a sufficiently long period of (proper) time. A particle that traverses this CTC will enter and exit in flat (chronology-respecting) spacetime, making it ideal for studying classical and quantum models of time travel \cite{friedman1990cauchy,echeverria1991billiard,tobar2020reversible,deutsch1991quantum,politzer1994path,lloyd2011closed,lloyd2011quantum,bishop2021time,bishop2022billiard}.

\begin{figure}
    \centering
    \includegraphics[scale=0.45]{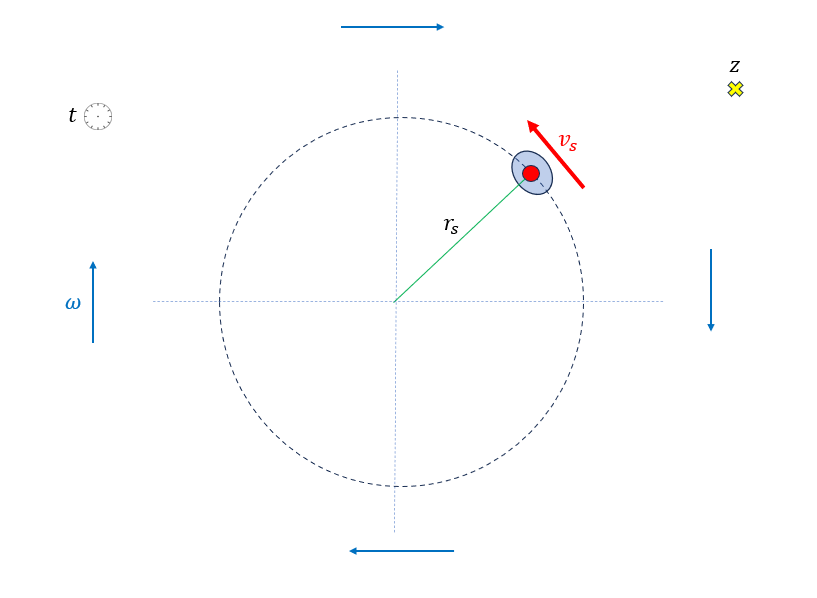}
    \caption{A diagram showing the basic features of the rotating Alcubierre spacetime. The Alcubierre bubble traverses a circular trajectory, and the entire setup is effectively placed on a disc that rotates at angular velocity $\omega$. \rev{The rotation of the disc induces a frame dragging effect inside the bubble. For simplicity, this effect can locally be reinterpeted as a modification to the bubble itself, while leaving the rest of the spacetime flat.}}
    \label{fig:rotating alc diag}
\end{figure}

%==========================================================================
\section{Modifying the original metric}
\label{sec:modify}

To be a useful background for quantum models of time travel, there should be at least one (ideally simple) classical path in the spacetime in which the particle definitively time travels. Although the path found by Ralph and Chang \cite{ralph2020spinning} (see Eq. \eqref{eq:four vel global}) does accomplish this, we find that a particle moving in such a trajectory would experience a nonzero radial acceleration,
\begin{equation}\label{eq:rad accel}
    a^\mu=u^\nu \nabla_\nu u^\mu=\left( 0,0,-\frac{\omega(\omega r_s+2v_s)}{1-\omega^2 r_s^2},0 \right).
\end{equation}
In obtaining this expression, we have assumed that the particle is initially contained in the $z=0$ hyperplane \rev{with $dz/d\tau=0$}, and $\partial_z f=0$ for all points near this hyperplane. \rev{The latter assumption ensures that the motion of a free particle will be constrained to the $z=0$ hyperplane, provided that the former initial conditions are satisfied.}

Notice that for $\omega\neq 0$, the acceleration \eqref{eq:rad accel} vanishes if and only if 
$$ v_s\omega r_s = -\frac{\omega^2 r_s^2}{2}>-\frac{1}{2}, $$
as the four-velocity components \eqref{eq:four vel global} must be real in order for the trajectory to be physical. This forbids the trajectory from being past-pointing, for which we require $v_s\omega r_s<-1$. Therefore, although the trajectory found by Ralph and Chang \cite{ralph2020spinning} is indeed a CTC, it fails to be a geodesic. In principle, one could consider the particle to be subject to some external classical field that keeps it confined to the trajectory. However, we expect that this would only result in further complications, such as the consideration of the how the field would be affected by the curved warp region around the bubble, as well as the backreaction of the field on the bubble itself. To avoid such difficulties, we shall restrict to the free case, with the aim of finding geodesics that are also CTCs. A recent study \cite{maclaurin2024falling} has shown that such paths exist in an eternal rotating Alcubierre spacetime, by replacing the localised bubble with a toroidal region in which $f=1$. However, the paths found in Ref. \cite{maclaurin2024falling} are complicated, and can only be solved for numerically. We will take an alternative approach to the problem by retaining the idea of a controlled, localised bubble region, but with modifications to the `rotation' of the spacetime which will be discussed in this section, aiming to find a simple CTC geodesic in the spacetime with closed form four-velocity components.

Notice that the radial acceleration \eqref{eq:rad accel} is a purely rotational effect, as it persists even if $v_s=0$. Ralph and Chang \cite{ralph2020spinning} treated the angular velocity $\omega$ as a uniform constant, originating from the idea of a Born-rigid rotating platform, although in principle this restriction is unnecessary. We thus make a slight modification to the metric, so that the $\omega=\omega(r)$ is radially dependent. In this case, we still find that $a^t=a^\phi=a^z=0$, while
\begin{equation}
    a^r=-\frac{\omega r_s(\omega+\omega' r_s)+v_s(2\omega+\omega'r_s(1+\omega^2 r_s^2))}{1-\omega^2 r_s^2},
\end{equation}
where $\omega'(r)=\frac{d\omega(r)}{dr}$. In order for this to be a geodesic, we require that $a^r=0$, which can be rearranged as
\begin{equation}\label{eq:v_s in terms of omega}
    v_s=-\frac{\omega^2 r_s+\omega\omega' r_s^2}{2\omega+\omega'r_s+\omega^2\omega' r_s^3}.
\end{equation}
We claim that this allows CTC trajectories that are also geodesics. For now, let us focus on the simplified case in which $v_s$ is constant. In this case, the trajectory is a CTC if and only if $v_s\omega r_s=-1$ (see Eq. \eqref{eq:four vel global}). For example, suppose $r_s=1$, $\omega(1)=-1/2$ and $v_s=2$. By Eq. \eqref{eq:v_s in terms of omega}, this CTC can only be a geodesic if $\omega'(1)=\frac{7}{8}$. One local choice of $\omega(r)$ that satisfies these conditions is
\begin{equation}\label{eq:omega prescription}
    \omega(r)=\frac{7}{8}r-\frac{11}{8}.
\end{equation}
Thus, this simple modification to the metric results in a spacetime that permits geodesic paths that are also CTCs. Hereafter, we shall refer to such paths as \emph{closed timelike geodesics} (CTGs).

Although our modification to the metric is successful in producing a simple CTG, a particle traversing this trajectory would always have $u^t=0$. This means that the particle would be stuck in an infinite time loop, never returning to chronology-respecting (flat) space within finite proper time intervals. A more physically meaningful situation to imagine is one in which the particle enters from flat space, interacts with a CTG, and eventually exits to flat space. Such a CTG would be ideal for studying situations in which a particle interacts with its past or future self, such as the billiard ball problem \cite{friedman1990cauchy,lossev1992jinn,echeverria1991billiard,novikov1989analysis,mikheeva1993inelastic,dolansky2010billiard}, and may generally serve as a useful playground for investigating classical and quantum models of time travel. In the next section, we will explain how such CTGs can be obtained.

%=========================================================================

\rev{\section{Closed timelike geodesics with a time-dependent rotation}\label{sec:CTG sudden}

\subsection{Formulating the problem}\label{subsec:formulating the problem}

In the previous section, we showed that the spacetime can be modified by introducing a radial dependence to $\omega=\omega(r)$,\footnote{\rev{In particular, this acts as an extra degree of freedom to control the frame dragging effect inside the bubble. The spacetime geometry outside the bubble remains flat.}} such that the circular bubble trajectories \eqref{eq:four vel global} found by Ralph and Chang \cite{ralph2020spinning} become geodesics for constant $v_s\neq 0$. Furthermore, this path can be made a CTC by choosing $v_s\omega(r_s)r_s=-1$. The final property that we desire of a time travel path is that a particle on this path enters and exits in flat spacetime. Since the other two properties are already satisfied, the most straightforward approach is to modify the warp drive to be dynamical, such that it accelerates up to some maximum value for which the bubble trajectory \eqref{eq:four vel global} is past-pointing relative to the coordinate time $t$, and eventually decelerates back to $v_s=0$. Indeed, as demonstrated by Ralph and Chang \cite{ralph2020spinning}, this can result in CTCs, on which a particle would enter and exit in flat space. However, this requires $v_s$ to depend on the proper time along the trajectory, which for one round trip from $0$ to $2\pi$ is equal to the local time\footnote{\rev{It should be noted that the local time, $t'$, is only equal to the proper time on the bubble trajectory for \textit{one} round trip, as $\phi=0$ and $\phi=2\pi$ are periodically identified. This is due to the lack of simultaneity in the Lorentz boost.}}
\begin{equation}
    t'=\frac{t-\omega(r_s) r_s^2 \phi}{\sqrt{1-\omega(r_s)^2 r_s^2}}.
\end{equation}
In such a case, $v_s$ would acquire a non-trivial dependence on $t$ and $\phi$, resulting in the bubble trajectory no longer being a geodesic.

Another natural idea would be to make the approximation that $v_s$ varies slowly in proper time, such that, over short proper time intervals, $v_s$ is approximately constant. This would allow us to simply patch together the Ralph and Chang \cite{ralph2020spinning} trajectories, producing an approximate CTG. However, this approach, whilst elegant at a glance, does not produce useful quantitative results. This is because there is no precise limiting case in which an exact CTG is recovered from the approximation, meaning that any calculations involving long proper time intervals cannot be trusted.

Given that the most straightforward approach does not yield a simple solution, it is worth considering more general scenarios. For simplicity, we will restrict to the case of timelike paths that are spatially circular, $x^\mu(\tau)$, which by our definition satisfy $dr/d\tau=dz/d\tau=0$. Furthermore, we require that $v_s=0$ at $\tau=0$ and $\tau=\tau_f$ (so that the particle enters and exits in flat spacetime), and that $x^\mu(\tau)$ is a geodesic. Finally, we can check if $x^\mu(\tau)$ allows time travel by inspecting $u^t=dt/d\tau$.

We first determine the four-velocity of a general circular timelike trajectory in the spacetime, at arbitrary constant radius $r=r_s$. It is easiest to work in local coordinates. Let $v=r_s \frac{d\phi'}{dt'}$ be the tangential velocity of a particle on this trajectory in the local coordinates. It can easily be shown that the four-velocity components in the local coordinates are
\begin{equation}\label{eq:local four vel general timelike}
    u^{\mu'}=\left( 1-(v-v_s)^2 \right)^{-1/2}\left( 1,\frac{v}{r_s},0,0 \right),
\end{equation}
where we have used the same ordering convention for components as Eq. \eqref{eq:four vel local}. We have assumed that the bubble is controlled such that the particle always remains within its centre, where $f=1$.\footnote{\rev{There is no issue with making this assumption, as we are not concerned with the stress-energy tensor required to produce bubbles of arbitrary shape (see Section \ref{sec:rotalc}).}} Notice that these components are real if and only if $|v-v_s|<1$. This is to be expected, as a physical massive particle should not locally attain or exceed the speed of light.

Transforming these components to the lab coordinates yields
\begin{equation}\label{eq:general circular lab}
\begin{split}
u^\mu = &\left( 1-(v-v_s)^2 \right)^{-1/2}\\ &\hspace{0.1cm}\cdot\bigg( 
\frac{1+v\omega r_s}{\sqrt{1-\omega^2 r_s^2}},
\frac{v+\omega r_s}{r_s\sqrt{1-\omega^2 r_s^2}},0,0 
\bigg).
\end{split}
\end{equation}
Here, for brevity, we have used $\omega$ to denote $\omega(r_s)$, and we shall continue to do so unless otherwise specified. The proper time parameterising this trajectory is
\begin{equation}\label{eq:general proper time}
    d\tau = \frac{\sqrt{1-\omega^2 r_s^2}}{1+v\omega r_s}\sqrt{1-(v-v_s)^2} dt.
\end{equation}
Notice that if $v=v_s$ in Eq. \eqref{eq:general circular lab}, we recover Eq. \eqref{eq:four vel global} as expected. Moreover, for arbitrary $v$, observe that $dt/d\tau<0$ for $v\omega r_s <-1$. Thus, general timelike circular trajectories of this type can be modified to allow time travel, using a similar protocol to that described in Ralph and Chang \cite{ralph2020spinning}.

Even when $v_s$ is constant, the metric has no obvious isometries, as the support, $f$, of the bubble, depends non-trivially on $t$ and $\phi$. This makes it intractable to solve the geodesic equations analytically, unless we make some approximations or consider a limiting case. Another option is to somehow make $\omega$ dependent on $(t,\phi)$ in addition to the radial dependence that we introduced earlier. Intuitively, for constant $v_s$, introducing a radial gradient to $\omega$ resulted in a tidal force opposing that due to the motion of the bubble. For an accelerating bubble, the tidal force on the particle changes as a function of $t'$. Thus, if we modify the spacetime so that $\omega$ also changes with $t'$, we can set its radial gradient to vary so that the resulting tidal force effectively cancels that due to the dynamical motion of the bubble. We will describe such a modification in the following section, in which we find an exact closed timelike geodesic with initial conditions in flat spacetime.

\subsection{Falling into the past with a time-dependent rotation}\label{subsec:falling into the past sudden limit}

For simplicity, we make the choice,
\begin{equation}\label{eq:vs choice}
    v_s(\tau)=\frac{v_m}{2}\left( \tanh\left( k(\tau-\tau_1) \right)-\tanh\left( k(\tau-\tau_2) \right) \right),
\end{equation}
with $k\gg 1$ and $0<\tau_1<\tau_2$, so that the bubble is approximately switched off for $0<\tau<\tau_1$ and $\tau>\tau_2$, and switched on with $v_s=v_m$ during the interval $\tau\in [\tau_1,\tau_2]$. This ensures that the particle's entry and exit will be in flat spacetime. Here, $\tau$ denotes the proper time parameterising a trajectory inside the bubble. We will further explain this dependence below.

Firstly, we want to find $v$ for which the trajectory \eqref{eq:general circular lab} is a geodesic for $v_s=0$, before the bubble is switched on (during $0<\tau<\tau_1$ and $\tau>\tau_2$). Notice that the trajectory found by Ralph and Chang \cite{ralph2020spinning} with $v=v_s$ is not a geodesic of flat spacetime, as the four-velocity components for $v_s=0$ are
\begin{equation}
    u^\mu=\frac{1}{\sqrt{1-\omega^2 r_s^2}}\left( 1,\omega,0,0 \right).
\end{equation}
The geodesics in Minkowski space are straight lines. Thus, the only `circular' geodesics (by our definition) in flat spacetime are those of particles at rest in the inertial frame with respect to which the cylindrical coordinates are defined. Thus, we require that the four-velocity components are $u^\mu=(1,0,0,0)$ for $v_s=0$. This corresponds to substituting $v=v_s-\omega r_s$ into Eq. \eqref{eq:general circular lab}.

For general $v_s$ (not necessarily zero), substituting $v=v_s-\omega r_s$ into Eq. \eqref{eq:general circular lab} yields
\begin{equation}\label{eq:final four vel}
    u^\mu=\left( 1+\frac{v_s\omega r_s}{1-\omega^2 r_s^2}, \frac{v_s}{r_s(1-\omega^2 r_s^2)},0,0 \right).
\end{equation}
We already know that this trajectory is immediately a geodesic while the bubble is switched off. We should be able to exploit the degree of freedom in choosing $\omega(r)$ to fix the trajectory to also be a geodesic with $v_s=v_m$, for which the trajectory is past-pointing relative to $t$. To demonstrate this, let us consider the four-acceleration components of a particle on the trajectory \eqref{eq:final four vel}. We find that $a^t=a^\phi=0$ if and only if
\begin{equation}\label{eq:transport eqn vs}
    \omega(r_s) r_s^2 \partial_t v_s + \partial_\phi v_s = 0.
\end{equation}
This is a transport equation whose general solution is 
\begin{equation}\label{eq:transport eqn vs sol}
    v_s(t,\phi)=v_s\left( t-\omega(r_s) r_s^2 \phi \right).
\end{equation}
Substituting $v=v_s-\omega(r_s) r_s$ into Eq. \eqref{eq:local four vel general timelike}, we see that
\begin{equation}
    \frac{dt'}{d\tau}=\frac{1}{\sqrt{1-\omega^2(r_s) r_s^2}}.
\end{equation}
Since $\omega(r_s)$ and $r_s$ are constant, this can be integrated to give
\begin{equation}
    \tau=\sqrt{1-\omega^2(r_s) r_s^2}t'=t-\omega(r_s) r_s^2 \phi,
\end{equation}
which holds for $0\leq \phi<2\pi$. This means that Eq. \eqref{eq:transport eqn vs sol} is automatically satisfied by any choice of $v_s$ that depends on the proper time along our trajectory, which includes our choice \eqref{eq:vs choice}.

Thus, we immediately have $a^t=a^\phi=0$ at all proper times, without even needing to fix $\omega(r_s)$ and $\omega'(r_s)$. All that remains is the radial component. We find that $a^r=0$ for $v_s=v_m$ if and only if
\begin{equation}\label{eq:ar=0 condition sudden limit}
\begin{aligned}
    -&\omega \left( 1+\omega r_s(\omega r_s-2v_m) \right) \\
    &+\omega' r_s\left( -1+\omega r_s(v_m-\omega r_s) \right) = 0.
\end{aligned}
\end{equation}

We remind the reader that $\omega$ and $\omega'$ are used to denote $\omega(r_s)$ and $\omega'(r_s)$, respectively. The condition for time travel ($dt/d\tau<0$) for our new trajectory is
\begin{equation}
    v_m\omega r_s < -1+\omega^2 r_s^2,
\end{equation}
which simplifies to
\begin{equation}\label{eq:sudden limit time travel condition}
    v_m > \omega r_s - \frac{1}{\omega r_s}.
\end{equation}

Let us now consider an example. Set $\omega=-1/2$ and $r_s=1$. In order to have time travel, we require that Eq. \eqref{eq:sudden limit time travel condition} is satisfied,
\begin{equation}
    v_m > \omega r_s - \frac{1}{\omega r_s}=-\frac{1}{2}+2=\frac{3}{2}.
\end{equation}
It therefore suffices to set $v_m=2$. In this case, the condition \eqref{eq:ar=0 condition sudden limit} for $a^r=0$ reduces to
\begin{equation}
    \omega'(r_s)=\frac{13}{18}.
\end{equation}
Now that we have the values of $\omega(r_s)$ and $\omega'(r_s)$, we can extend this linearly to the rest of the bubble as
\begin{equation}\label{eq:omega_r_choice}
    \omega(r)=\frac{13}{18}r-\frac{11}{9}.
\end{equation}

Thus, for the chosen values of $v_m$, $\omega(r)$, and $r_s$, the four-acceleration components for the trajectory vanish at almost all proper times since $k\gg 1$ (see Eq. \eqref{eq:vs choice}). However for the trajectory to be a geodesic, the acceleration must be zero at all proper times, even as the bubble accelerates from $v_s=0$ to $v_s=v_m$.

One idea is to modify the metric so that $\omega$ depends on $\tau=t-\omega(r_s)^2 r_s$. We want to do this in such a way that our trajectory still remains valid (notice that $\tau$ itself depends on $\omega$ at $r=r_s$). It is then natural to choose $\omega$ so that it remains constant at $r=r_s$ for all $\tau$, while its radial gradient at $r=r_s$ varies with $\tau$. Only the radial gradient of $\omega$ needs to vary with $\tau$ to balance the changing `force' due to the motion of the bubble. We thus make the ansatz
\begin{equation}\label{eq:omegatau_ansatz}
    \omega(\tau,r)=\omega_0(r)+(r-r_s)\alpha(\tau),
\end{equation}
where $\alpha$ is some function of $\tau$ that remains to be specified, and $\omega_0$ is a function of $r$. For our previous example, we would choose $\omega_0(r)=\frac{13}{18}r-\frac{11}{9}$ (see Eq. \eqref{eq:omega_r_choice}). Given the ansatz \eqref{eq:omegatau_ansatz}, we have
\begin{align}
    \partial_\tau \omega(\tau,r)&=(r-r_s)\partial_\tau \alpha(\tau), \\
    \partial_r \omega(\tau,r) &= \partial_r \omega_0(r)+\alpha(\tau).
\end{align}
Physically, $\omega$ can loosely be interpreted as describing the velocity of a row of conveyor belts inside the bubble, as described in Section \ref{sec:rotatingalc}. Thus, in the modified metric, the bubble is deformed so that the velocity of these conveyor belts is dependent on their spacetime location, except for the belt at $r=r_s$. To see this, notice that $\partial_\tau \omega(\tau,r_s)=0$ and $\partial_r \omega(\tau,r_s)=\omega'_0(r_s)+\alpha(\tau)$. In other words, $r=r_s$ is a fixed point of $\omega$. Importantly, the $t$ and $\phi$ four-acceleration components corresponding to our trajectory \eqref{eq:final four vel} still vanish, $a^t=a^\phi=0$, even with this modification.

For simplicity, we choose units in which $r_s=1$. We can now fix $\alpha(\tau)$ for any given $v_s(\tau)$ by requiring that $a^r=0$ for all $\tau$ as follows:
\begin{equation}
    \alpha(\tau)=-\frac{\omega_0 \left( 1-2v_s(\tau)\omega_0+\omega_0^2 \right)}{1-v_s(\tau)\omega_0+\omega_0^2}-\omega_0'.
\end{equation}
Here, $\omega_0$ and $\omega_0'$ are used to denote $\omega_0(1)$ and $\omega_0'(1)$ for brevity. For our earlier example, this reduces to 
\begin{equation}
    \alpha(\tau)=\frac{5}{9}\frac{v_s(\tau)-2}{2v_s(\tau)+5}.
\end{equation}
We remind the reader that $\tau$ can be written as a linear combination of $t$ and $\phi$, as $\tau=t-\omega_0(r_s) r_s^2\phi$.

Finally, we want to choose $\tau_1$, $\tau_2$, and $\tau_f$ such that our trajectory is a CTC. To elaborate, we require that for some $\tau_f>\tau_2$, $t_f\equiv t(\tau_f)=0$ and $\phi_f\equiv \phi(\tau_f)=2\pi$. We choose initial conditions $t(\tau=0)=\phi(\tau=0)=0$ so that the particle enters and exits in flat space. Using the four-velocity components \eqref{eq:final four vel}, we require that the following equations are satisfied:
\begin{align}
    2\pi r_s &= \frac{1}{1-\omega^2 r_s^2}v_m(\tau_2-\tau_1) \\
    0 &= \tau_f+\frac{\omega r_s}{1-\omega^2 r_s^2}v_m(\tau_2-\tau_1).
\end{align}
To derive this system of equations, we have assumed that $k\gg 1$ so that the integral of $v_s(\tau)$ is easy to approximate (see Eq. \eqref{eq:vs choice}). Solving this system for $\tau_f$ and $\tau_2-\tau_1$ gives
\begin{align}
    \tau_f &= -2\pi \omega r_s^2 \\
    \tau_2-\tau_1 &= \frac{2\pi r_s(1-\omega^2 r_s^2)}{v_m}.
\end{align}
For the chosen values of $\omega$, $r_s$ and $v_m$ in our example, we have $\tau_f=\pi$. If we now choose $\tau_1=\frac{1}{2}$, we have
\begin{equation}
    \tau_2 = \frac{1}{2}+\frac{3\pi}{4} < \pi = \tau_f.
\end{equation}
Thus, the trajectory with these chosen values of the free parameters is a CTG.}

%==========================================================================
\rev{\section{Discussion}\label{sec:Discussion}

In the previous sections, we have shown that the rotating Alcubierre metric, originally due to Ralph and Chang \cite{ralph2020spinning}, can be modified to allow for simple CTGs by introducing a time-dependent internal flow to the bubble. In this section, we will further discuss the physical meaning of our model.

In Section \ref{subsec:global coord dep of bubble vel}, we will explain why the proper time dependence of the warp velocity $v_s=v_s(\tau)$ is not problematic. It seems that the $\tau$-dependence implies that the bubble has to be `controlled' by a local agent inside the bubble. However, such a scenario is causally forbidden for superluminal warp drives \cite{krasnikov1998hyperfast}, as there exists an outer white hole horizon near the front wall of the bubble \cite{alcubierre1994warp}. Instead, we will show that $v_s$ can equivalently be written as a function of the global coordinates $(t,\phi)$, allowing the warp field to be controlled by a global agent situated outside the bubble. This is similar to the argument used by Krasnikov \cite{krasnikov1998hyperfast} to demonstrate that the steering of a superluminal Alcubierre bubble does not necessarily require tachyons. Instead, an agent, whose future lightcone contains the trajectory of the bubble, could in principle prearrange devices along the desired trajectory and program them to `switch on' and create the bubble at preassigned times \cite{everett1997superluminal}. Strictly speaking, such a setup requires a relaxation of the condition that the spacetime is initially flat, as the devices themselves would have non-vanishing stress-energy tensor.

In Section \ref{subsec:periodically identified alcubierre}, we discuss a different (topologically nontrivial) time travel metric obtained by periodically identifying the Alcubierre metric. To our knowledge, this simple modification has not yet been presented in the literature. We argue that it might be possible to formulate quantum field theory on this metric, making it ideal for studying quantum effects near CTCs.

\subsection{Global coordinate dependence of the bubble velocity}\label{subsec:global coord dep of bubble vel}

Let us phrase the problem of controlling the bubble in an operational sense. Suppose hypothetically that there existed a machine that could locally manipulate the curvature of spacetime in any way we desired. Assume that the stress-energy tensor of the machine itself does not result in a significant backreaction on the spacetime. How could we use such a machine (or machines) to realise the rotating Alcubierre metric with $v_s(\tau)$ as in Eq. \eqref{eq:vs choice}?

As mentioned earlier, the warp field cannot be controlled by a local agent inside the bubble, as this would require a local violation of causality, which conflicts with special relativity. To resolve this apparent issue, we note that the proper time along our trajectory can be written in terms of the global coordinates for $0\leq \phi < 2\pi$:
\begin{equation}
    \tau=t-\omega_0(r_s) r_s^2 \phi,
\end{equation}
as we showed in Section \ref{subsec:falling into the past sudden limit}. Thus, in terms of the global coordinates,
\begin{equation}\label{eq:vs global description sudden limit}
\begin{split}
    v_s(t,\phi) = \frac{v_m}{2} \Big[ \tanh\Big( k(t 
    - \omega_0(r_s) r_s^2 \phi - \tau_1) \Big) \\ - \tanh\Big( k(t - \omega_0(r_s) r_s^2 \phi - \tau_2) \Big) \Big],
\end{split}
\end{equation}
for $(t,\phi,r_s,0) \in \supp{f}$.

\begin{figure}
    \centering
    \includegraphics[width=1\linewidth]{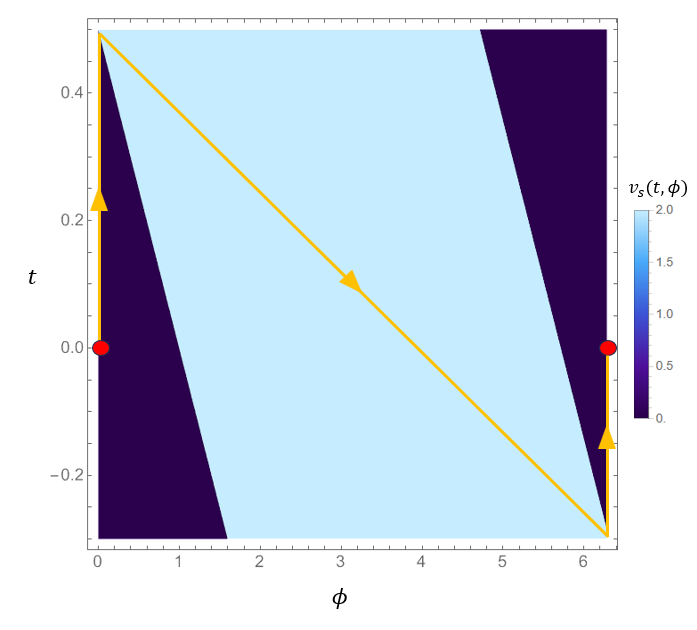}
    \caption{\rev{A density plot of $v_s$ as a function of the global coordinates, $(t,\phi)$, in the sudden limit. Here, we have chosen the parameters $r_s=1$, $\omega=-1/2$, $\tau_1=1/2$, $\tau_2=1/2+3\pi/4$, and $\tau_f=\pi$. The points marked in red are the initial and final points of the particle's trajectory, respectively $(t,\phi)=(0,0)$ and $(t,\phi)=(0,2\pi)$. These points are the same since $\phi=0$ and $\phi=2\pi$ are periodically identified. The orange line plotted over $v_s$ is the worldline of the particle, corresponding to the integral curve of Eq. \eqref{eq:final four vel}, with the arrows indicating the direction of increasing $\tau$.}}
    \label{fig:sudden regime vs in global coords}
\end{figure}

For the example considered in the previous section, a plot of $v_s(t,\phi)$ is shown in Fig. \ref{fig:sudden regime vs in global coords}. Thus, assuming the existence of the aforementioned spacetime-manipulating machines,\footnote{\rev{Of course, such technology does not currently exist.}} our metric could be realised by placing multiple machines around a circle at $r=r_s$. Then, each of them could be switched on during different global time intervals, and switched off at all other times. For each machine at a fixed $\phi$, this time interval is determined using Eq. \eqref{eq:vs global description sudden limit}, visually represented in Fig. \ref{fig:sudden regime vs in global coords}. The centre ($f=1$) of the bubble generated by the machines would correspond to a tube within the support of $v_s$ around the worldline of the particle, represented by the orange line in Fig. \ref{fig:sudden regime vs in global coords}. Assuming that the huge amounts of exotic matter required to produce the bubble could be realised, our time travel metric could also be realised without requiring local violations of causality.

\subsection{A periodically identified Alcubierre spacetime}\label{subsec:periodically identified alcubierre}

Consider the original rotating Alcubierre metric, as formulated by Ralph and Chang \cite{ralph2020spinning} in the lab coordinates. We can then take the induced metric on the $r=r_s$ hypersurface, and re-embed this metric in the following $(3+1)$-dimensional spacetime:
\begin{align}\label{eq:rectangular line element}
    ds^2 &= -\frac{1-(v_r+fv_s)^2}{1-v_r^2}\: dt^2\nonumber\\&\hspace{0.4cm}-\frac{2(fv_s+f^2 v_s^2 v_r+fv_s v_r^2)}{1-v_r^2}\: dx\: dt\nonumber\\&\hspace{0.4cm}+\frac{(1+fv_s v_r)^2-v_r^2}{1-v_r^2}\: dx^2+dy^2+dz^2,
\end{align}
where $y$ is the rectangular coordinate in the direction of the added spatial dimension. Here, we have defined $x\equiv r_s\phi$ and $v_r\equiv \omega r_s$. One nice property of this modified metric is that the original circular bubble trajectory \eqref{eq:four vel global} is immediately a geodesic for any choice of $v_s$ that depends on the spacetime coordinates. This is to be expected as the spacetime described by the metric \eqref{eq:rectangular line element} is equivalent to the Alcubierre spacetime \cite{alcubierre1994warp} with the topological identification $x \sim x+2\pi r_s$. One can see this by performing a Lorentz boost in the $x$ direction by velocity $v_r$, from which the line element \eqref{eq:alcubierre} can be recovered. Because of the periodic identification, the bubble trajectory can also be made a CTC, using a similar procedure to that employed by Ralph and Chang \cite{ralph2020spinning}. Thus, this trivial modification to the Alcubierre metric (which is globally hyperbolic) contains simple CTGs. Conversely, the Ralph and Chang metric can be viewed as a cylindrical version of the boosted Alcubierre metric. However, the spacetime described by Eq. \eqref{eq:rectangular line element} is not Minkowski outside the compact bubble region---unlike the rotating Alcubierre spacetime---due to the topological identification.\footnote{\rev{The spacetime outside the bubble region still has vanishing Riemann curvature and is chronology-respecting.}}

Given that the original Alcubierre metric is globally hyperbolic, it should be possible to formulate quantum field theory (QFT) on it using known techniques. Then, it should also be possible to formulate QFT on the periodically identified Alcubierre metric using the formalism of automorphic fields \cite{langlois2006causal}. In this formalism, the quantum fields are first studied on the universal covering space (in our case, the original Alcubierre spacetime), and imposing certain conditions on them (such as periodic boundary conditions) \cite{banach1979vacuum,banach1980quantum,frolov1991vacuum}. Such an approach was recently employed in Refs. \cite{alonso2021time} and \cite{alonso2024particle} for a different time machine spacetime, constructed by a topological identification of the Po\'incare patch of two-dimensional Anti-de Sitter ($\text{AdS}_2$) spacetime, which itself is not globally hyperbolic. Thus, in principle, it should be possible to formulate QFT on the periodically identified Alcubierre metric \eqref{eq:rectangular line element}, suggesting that this spacetime might be useful for studying the behaviour of quantum fields near CTCs that are embedded in chronology respecting space.\footnote{It should be noted that although the Alcubierre spacetime is globally hyperbolic, no full treatment of QFT has been formulated on it as far as we are aware. However, calculations of quantum effects near an Alcubierre bubble have been considered, particularly to do with Hawking radiation and semiclassical instability of the metric under backreaction effects \cite{hiscock1997quantum,finazzi2009semiclassical,jusufi2018quantum} using the tunnelling method (see Ref. \cite{vanzo2011tunnelling} for a review). It might be possible to relate these calculations in the tunnelling formalism to a standard QFT treatment \cite{helfer2019hawking}.} We suggest this as a possible direction for future work.

Cosmological observations strongly suggest that we likely do not live in a closed universe \cite{aghanim2020planck}, meaning that the periodically identified Alcubierre metric loses the clear physical interpretation characteristic of the rotating Alcubierre metric. However, given that this spacetime still contains CTCs embedded in an otherwise chronology-respecting space (outside the bubble), it may still serve as a useful toy model for investigating quantum effects near CTCs.}
%==========================================================================

\section{Conclusions}
\label{sec:conclusions}

In this paper, we have given a modification to the time travel metric due to Ralph and Chang \cite{ralph2020spinning}. \rev{The Ralph and Chang metric was first obtained by considering an Alcubierre bubble in orbit around a Kerr black hole.\footnote{\rev{This was confirmed through private communication with T.C. Ralph and C. Chang.}} In particular, the rotation of the black hole induces a frame dragging effect inside the bubble. This dragging effect is what leads to the possibility of time travel. Mathematically, it was shown by Ralph and Chang that this is locally equivalent to a modification to the warp bubble itself. This results in a simplification in which the Kerr black hole can be removed and the spacetime is flat outside the compact bubble region, while preserving the desirable time travel properties of the metric. The remaining `rotational' effect that is confined to the bubble can then loosely be viewed as being due to a series of massive conveyor belts inside the bubble, or the internal flow of a fluid confined to the bubble.

Although the path found by Ralph and Chang is a CTC with initial conditions in flat (chronology-respecting) spacetime, we showed that it is not a geodesic, as a nonzero radial acceleration is required. Physically, confining a particle to this CTC requires the introduction of an external field. It is difficult to determine how such a field would be affected by the warp region. Furthermore, the backreaction of the field on the bubble would have to be negligible. To avoid such difficulties, we instead took the approach of modifying the metric so that the tidal forces due to the motion of the bubble would be balanced on a particular trajectory. Specifically, we introduced an ad hoc dependence of $\omega$ on the spacetime coordinates, rather than demanding that it be uniform throughout the bubble. Physically, this corresponds to the aforementioned conveyor belts changing velocity as the warp bubble accelerates. We have described a scenario in which the chosen trajectory is both a CTC and a geodesic, allowing the free fall of classical pointlike particles into the past. 

One problem that we have not addressed in this paper is the violation of energy conditions, and whether our spacetime could be realised through quantum effects. Given that our time machine spacetime construction is based on an Alcubierre bubble, it is reasonable to expect that the classical energy conditions are still violated. It would be interesting to study whether the structure of the bubble can be modified---as was done for the original Alcubierre metric in Refs. \cite{pfenning1998quantum,van1999awarp,krasnikov2003quantum}---so that QEIs are satisfied. There is also the question of whether the semiclassical instabilities of the original Alcubierre metric \cite{hiscock1997quantum,finazzi2009semiclassical} also appear in our metric. There is no immediate reason to expect that our modified spacetime should exhibit these instabilities, as it was shown in Ref. \cite{barcelo2022warp} that Alcubierre bubbles on certain `zig-zag' trajectories should remain stable, demonstrating that not all modifications of the Alcubierre metric are semiclassically unstable. We leave the consideration of these issues to future work.

Our spacetime contains similar desirable features to the idealised traversable wormhole metric, often employed in time travel models, without neglecting any aspects of the geometrical structure while (for the most part) retaining mathematical simplicity. We have also shown that it can be related to the even simpler periodically identified Alcubierre spacetime, which also contains CTGs with initial conditions in (periodically identified) flat spacetime. We have argued that it should be possible to formulate QFT on this metric using the formalism of automorphic fields, by first formulating QFT on the original (unidentified) Alcubierre metric. We expect that such a study, if manageable, would lead to better physical insight into the compatibility of quantum time travel models with general relativity. Ultimately, this may lead to physical restrictions on these models beyond the level of mathematical self-consistency, which could facilitate a better understanding of quantum effects near CTCs.}

%==========================================================================
\begin{acknowledgments}
    We thank Nicolas C. Menicucci, Nicholas Funai, Eduardo Mart\'in-Mart\'inez, Colin Maclaurin, and Alex Tremayne for useful discussions. This research was supported by the Australian Research Council Centre of Excellence for Quantum Computation and Communication Technology (Project No. CE170100012).
\end{acknowledgments}

%%%%%%%%%%%%%%%%%%%%%%%%%%%%%%%%%%%%%%%%%%%%%%%%%%%%%%%%%%%%%%%%%%%%%%%%%%%%%%%
%\section{References}
%%%%%%%%%%%%%%%%%%%%%%%%%%%%%%%%%%%%%%%%%%%%%%%%%%%%%%%%%%%%%%%%%%%%%%%%%%%%%%%
\bibliography{ref}

\end{document}